\documentclass[prb,superscriptaddress,showpacs,floatfix,preprint]{revtex4-1}

\usepackage{graphicx}
\usepackage[graphicx]{realboxes}
\usepackage{amsmath}
\usepackage[english]{babel}
\usepackage{enumerate}
\usepackage{tabularx}
\usepackage{multirow}
\usepackage{array}
\usepackage{float}
\usepackage[caption=false]{subfig}
\newcolumntype{C}[1]{>{\centering\let\newline\\\arraybackslash\hspace{0pt}}m{#1}}

\usepackage[update,prepend]{epstopdf}
\begin{document}

%%%%%%%%%%%%%
%\linenumbers
%%%%%%%%%%%%%

\bibliographystyle{apsrev4-1}

\title{Micromagnetic study of skyrmion stability in confined magnetic structures with perpendicular anisotropy}
%\title{Static and dynamic skyrmion stability in confined magnetic structures with perpendicular anisotropy}
%\today

\author{R. L. Novak}
\email[Electronic address: ]{rafael.novak@ufsc.br}
\affiliation{Universidade Federal de Santa Catarina, Campus Blumenau, Rua Pomerode, 710 Blumenau 89065-300, SC, Brazil.}
\author{F. Garcia}
\affiliation{Centro Brasileiro de Pesquisas F\'isicas -- CBPF, Rua Dr. Xavier Sigaud, 150 Rio de Janeiro 22180-790, RJ, Brazil} 
\author{E. R. P. Novais}
\affiliation{Faculdade de F\'isica -- Universidade Federal do Sul e Sudeste do Par\'a, Marab\'a, PA 68500-970, PA, Brazil} 
\author{J. P. Sinnecker}
\author{A. P. Guimar\~aes}
\affiliation{Centro Brasileiro de Pesquisas F\'isicas -- CBPF, Rua Dr. Xavier Sigaud, 150 Rio de Janeiro 22180-790, RJ, Brazil}

%%%%%%% ABSTRACT %%%%%%%%%
\begin{abstract}

%%% NEW abstract, better! Sent to SBPMat 2017 %%%% Replace previous by this one!
Skyrmions are emerging topological spin structures that are potentially revolutionary for future data storage and spintronics applications. The existence and stability of skyrmions in magnetic materials is usually associated to the presence of the Dzyaloshinskii-Moriya interaction (DMI) in bulk magnets or in magnetic thin films lacking inversion symmetry. While some methods have already been proposed to generate isolated skyrmions in thin films with DMI, a thorough study of the conditions under which the skyrmions will remain stable in order to be manipulated in an integrated spintronic device are still an open problem. The stability of such structures is believed to be a result of ideal combinations of perpendicular magnetic anisotropy (PMA), DMI and the interplay between geometry and magnetostatics. In the present work we show some micromagnetic results supporting previous experimental observations of magnetic skyrmions in spin-valve stacks with a wide range of DMI values. Using micromagnetic simulations of cobalt-based disks, we obtain the magnetic ground state configuration for several values of PMA, DMI and geometric parameters. Skyrmion numbers, corresponding to the topological charge, are calculated in all cases and confirm the occurrence of isolated, stable, axially symmetric skyrmions for several combinations of DMI and anisotropy constant. The stability of the skyrmions in disks is then investigated under magnetic field and spin-polarized current, in finite temperature, highlighting the limits of applicability of these spin textures in spintronic devices.

\end{abstract}

%75.60.-d 	Domain effects, magnetization curves, and hysteresis
%75.60.Ch Domain walls and domain structure (for magnetic bubbles, see 75.70.Kw)
%75.78.Fg	Dynamics of domain structures

%%%%%%%%%%%%%%%%%%%%%%% MAIN BODY %%%%%%%%%%%%%%%%%%%%%%%
\maketitle

\section{Introduction \label{intro}}

Skyrmions are small size topological structures with very good mobility under spin-polarized currents that are being considered for use in future magnetic memories and devices in the emergent field already known as ``Skyrmionics''\cite{Krause2016,nature_editorial,nagaosa2013,fertcros2013}. Their stability is a major issue regarding their application in devices, especially at room temperature, where knowledge of the conditions for skyrmion stability in the absence of external excitations (zero magnetic field, zero spin polarized electric current) is fundamental for applications in non-volatile data storage devices\cite{beach_preprint,Hagemeister2015,beg_scirep}. The existence of such spin structures is usually associated to the presence of the Dzyaloshinskii-Moriya interaction (DMI) in bulk magnets lacking inversion symmetry or in thin film structures grown on suitable substrates\cite{rossler2006,muhlbauer2009} where the lack of translational symmetry across the interface (or across different layers in multilayer stacks), combined with the presence of heavy atoms with significant spin-orbit interaction in the substrate, give rise to a sizable DMI\cite{heinze2011}. In analogy to other topological structures in magnetism, skyrmions are characterized by a topological charge related to the winding number, also known as the skyrmion number\cite{nagaosa2013,Tretiakov2007}, which can assume the following values: $Sk = +1$ or $-1$, where the negative value corresponds to an anti-skyrmion. Practical applications of skyrmions remain a challenge, because in most cases they can only be observed in a limited choice of materials, generally at low temperatures and within narrow ranges of applied magnetic field.

Several methods have been proposed to generate isolated skyrmions in magnetic nanostructures with DMI, based both on micromagnetic simulations and on experimental results: a suitable combination of electrical current and applied magnetic field could drive skyrmions out of the edge of notches in nanowires\cite{iwasaki2013} or from the edges of nanodisks\cite{sampaiocros2013}; local heating by focused laser light can locally induce spin flipping, producing a skyrmion\cite{koshibae2014}; overlying nanodisks in a vortex state can induce the formation of skyrmions in a magnetic film\cite{sun2013,loreto2017}; suitable combinations of perpendicular magnetic anisotropy constant, disk radius and disk thickness led to skyrmionic ground states in the absence of DMI \cite{Novais2011}; and domain wall pairs, driven by spin-polarized currents from a narrow nanowire towards a broad nanowire, continuously transform into skyrmions at the border between them\cite{zhouezawa2014}. Further experimental demonstrations of the feasibility of creating single skyrmions are found in Romming et al.\cite{romming2013}, where local spin-polarized currents from a STM tip are used to write and erase skyrmions in magnetic thin films at low temperatures; in ref. \cite{boulle_nature}, where the authors experimentally observed a stable skyrmion at room temperature in the absence of magnetic fields; in refs.\cite{beach_preprint} and \cite{fert_preprint}; and in ref. \cite{jiang_science}, where stripe domains driven through constrictions by inhomogeneous currents are converted into skyrmions. Despite these efforts, an unequivocal, general-purpose way to generate and stabilize skyrmions in magnetic nanostructures without stringent requirements on temperature, applied current, DMI and magnetic anisotropy, is still lacking.

In this work we show that by judiciously tailoring the perpendicular magnetic anisotropy (PMA) and the Dzyaloshinskii-Moriya interaction (DMI) in Co-based magnetic disks, it is possible to stabilize skyrmions for certain combinations of these material parameters. In certain cases, the skyrmion may be stable even in the absence of DMI. In order to describe the magnetic configuration of similar samples and to find out the limits of applicability of the skyrmions eventually found within suitable ranges of DMI and PMA, we have analyzed micromagnetic simulations of Co films with the appropriate geometries and under the influence of different excitations: magnetic field, electric spin-polarized current and temperature. The skyrmion numbers are calculated in all final states of the simulated structures, confirming the existence of isolated, stable N\'eel type (hedgehog, \emph{HG}) or Bloch type (vortex-like, \emph{VL}) skyrmions\footnote{We stick to the nomenclature used by Sampaio et al. in \cite{sampaiocros2013}. In other works, an alternative nomenclature is used, where a vortex-like skyrmion is called a Bloch skyrmion, and a hedgehog skyrmion becomes a N\'eel skyrmion.} in nanosize disks under certain conditions. In the following, phase diagrams obtained for different initial magnetic configurations of the disks, along with studies of skyrmion stability at finite temperature, under applied magnetic fields and spin-polarized currents, will be presented and discussed.

%%%%%%%%%%%%%%%%%%%%%%%%%%%%%%%%%%%%%%%%

\section{Methods \label{sec:met}}

The computational studies presented were done using the \textsc{Mumax3} micromagnetic simulation package\cite{mumax3}. The simulated geometry consisted of a 2 nm thick disk with cobalt material parameters ($A_{ex} = 30$ pJ/m, $M_{sat} = 1400$ kA/m), 1 x 1 x 2 nm$^3$ simulation cells and $\alpha = 0.5$, for fast relaxation. The perpendicular uniaxial anisotropy constant ($K_{u}$) and the interface-induced Dzyaloshinskii-Moriya exchange constant, $D_{ex}$, were varied during the simulations in order to map combinations of these parameters that can lead to stable skyrmion ground states, with the probed ranges corresponding to typical experimental values found in the literature \cite{heinze2011,sampaiocros2013}.

The disks were set to one of these three magnetization states prior to each simulation: vortex-like (VL), hedgehog (HG) \cite{sampaiocros2013} or uniform perpendicular magnetization. Then, the nanodot magnetization is relaxed in the absence of applied magnetic fields or electrical currents, with the simulation temperature set to $0$ or to $300$ K. \textsc{Mumax3} can introduce finite temperatures in the simulations by means of a random field term in the LLG equation, given according to the Brown-Langevin fluctuating field \cite{mumax3,brown_thermal,brown_thermal2,PhysRevB.93.214412}. This procedure was repeated for values of $K_{u}$ comprised between $1.1$ MJ/m$^3$ and $1.4$ MJ/m$^3$, in 5 kJ/m$^3$ steps. For each value of $K_{u}$, the DMI exchange constant, $D_{ex}$, was varied from $0$ to $10$ mJ/m$^2$ in 0.2 mJ/m$^2$ steps, a range of values commonly found in the literature\cite{sampaiocros2013} for this kind of sample. After the disk reached its ground state, the corresponding skyrmion number is calculated according to a finite difference implementation\cite{nagaosa2013} of Eq. (\ref{eq:winding1}). The ground state is reached in each simulation either by means of a conjugate gradient energy minimization routine or by time evolution of the Landau-Lifshitz-Gilbert equation until the Landau-Lifshitz torque is inferior to $10^{-4}$ rad/s. No significant differences were observed in the results obtained from both methods. This method allowed us to investigate broad regions of the $K_u$ \emph{vs.} $D_{ex}$ parameter space for all geometries and initial conditions (VL, HG or uniform saturated) studied. In order to test the stability under magnetic fields or spin polarized currents, a few stable skyrmion configurations were submitted to uniform, out-of-plane currents or fields increasing in intensity until skyrmion annihilation is observed.

%%%%%%%%%% Equation 1 %%%%%%%%%%%%
\begin{equation}
Sk = \frac{1}{4 \pi} \int \!\!\! \int \vec m \cdot \big(\frac{\partial \vec m }{\partial x} \times \frac{\partial \vec m }{\partial y} \big) dx dy
\label{eq:winding1}
\end{equation}
%%%%%%%%%%%%%%%%%%%%%%%%%%%

\section{Results and Discussion \label{results}}

\subsection{Skyrmion stability at $T = 0$ K in the absence of magnetic fields and electric currents -- phase diagrams} \label{sec:diagrams}

\begin{figure}
\centering
\includegraphics[width=1\linewidth]{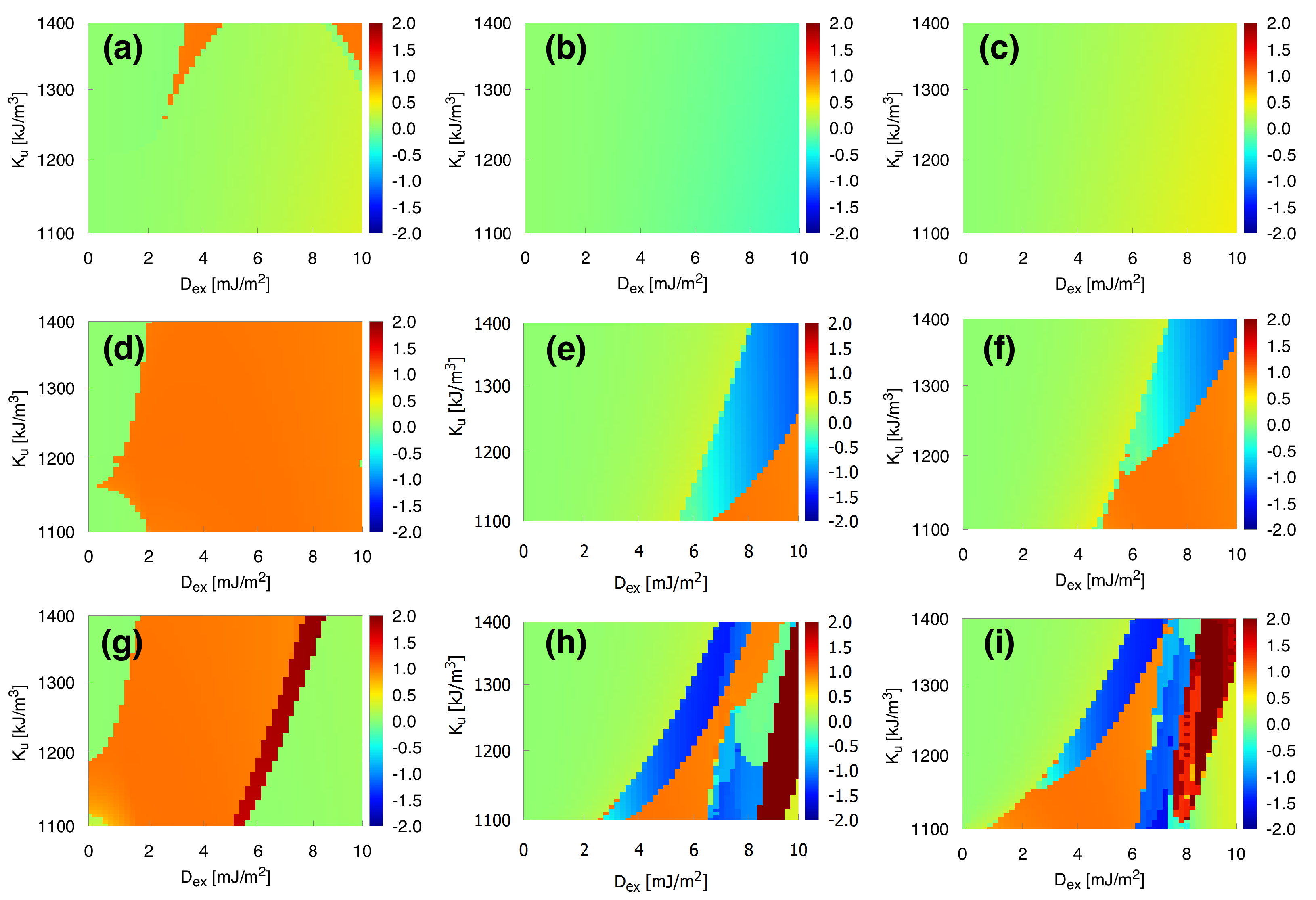}
\caption{Skyrmion number phase diagrams of disks with a skyrmion (VL or HG) or uniform initial state, calculated according to (\ref{eq:winding1}). Each diagram represents regions of the parameter space where $K_{u}$ lies in the $1100$ -- $1400$ kJ/m$^3$ range and $D_{ex}$ is comprised between $0$ and $10$ mJ/m$^2$. In all diagrams, the color indicates the skyrmion number \emph{Sk}, with orange representing the stable skyrmion ground state ($Sk = 1$). (a--c) $32$ nm disk with VL, HG and uniform initial states, respectively; (d--f) $64$ nm disk with VL, HG and uniform initial states, respectively; (g--i) $128$ nm disk with VL, HG and uniform initial states, respectively. (Color online.)}
\label{fig:maps_blochneel}
\end{figure}

This procedure allowed us to construct phase diagrams mapping skyrmion number to combinations of $K_{u}$ and $D_{ex}$ in the absence of magnetic field and electric currents, with $T = 0$ K. Regions with $Sk = 0$ correspond to uniform states or states with a single domain wall. Regions with $Sk = 1$ correspond to a single stable skyrmion in the disk. Regions with $|Sk| > 1$ correspond to more complex multidomain phases which will not be discussed. Diagrams for disks with different diameters ($32$, $64$ and $128$ nm diameter or side) and initial states (VL, HG or saturated) are shown in Fig. \ref{fig:maps_blochneel}.

%%%%%%%%%%%%%%% Figure 1 %%%%%%%%%%%%%%%
%%%%%%%%%%%%%%%%%%%%%%%%%%%%%%%%%%%%%%

The diagrams show that $32$ nm diameter disks only present a stable skyrmion ground state at 0 K for a few combinations of anisotropy constant and DMI, if the initial state is VL (the orange regions in Fig. \ref{fig:maps_blochneel}, diagram (a)). In the other scenarios tested (HG and uniform initial states), a skyrmion cannot be stabilized in the ground state within the ranges of anisotropy constant and DMI investigated. The phase diagrams of $64$ and $128$ nm disks show that a more complex picture emerges when the disk size is increased. Broad regions with $Sk = 1$ (Fig. \ref{fig:maps_blochneel}(d) -- (i), orange) indicate that a stable skyrmion ground state could be reached for many combinations of $K_{u}$ and $D_{ex}$ in these disks. The 64 nm diagrams display large regions where a stable skyrmion ground state was reached, with the region size strongly dependent on the initial state of the disk in the simulation (Fig. \ref{fig:maps_blochneel}(d) -- (f), orange). Disks initially in a VL state reached a stable skyrmion ground state for most combinations of $K_{u}$ and $D_{ex}$ studied, while for a HG initial state, the stability region is smaller and confined to a low $K_{u}$ and high $D_{ex}$ part of the diagram. Uniform initial states yielded a larger stability region, again for low $K_{u}$ and high $D_{ex}$.

In $128$ nm disks (Fig. \ref{fig:maps_blochneel} (g), (h) and (i)), more complex diagrams were obtained, with stable skyrmion ground states occurring for several combinations of $D_{ex}$ and $K_{u}$ (orange regions) and for all initial conditions.  In these diagrams, regions with $|Sk| > 1$ are present, being related to more complex multidomain structures stabilized when the simulation reaches equilibrium, as has already been reported by Sampaio et al.\cite{sampaiocros2013} and Zhou et al.\cite{PhysRevB.93.024415}. Values which are not multiples of $1/2$ are also present in the diagrams, being related to the discrete distribution of spins in the simulations. In theory, skyrmions possess an inherent topological protection, for it is theoretically not possible to continuously change the topological charge from $0$ to $1$, e.g., transforming a domain wall ($Sk = 0$) into a skyrmion\cite{nagaosa2013} ($Sk = \pm 1$). This fact is related to the well known difficulty to create skyrmions in magnetic thin films. This topological approach, however, is based on a continuous description of the magnetic crystal, while in real samples it breaks down at atomic length scales due to the discreteness of the magnetic crystal lattice, while in computational micromagnetic descriptions, the continuous approach breaks down due to the discretization of the simulated samples into finite difference cells (1 x 1 x 2 nm$^3$ cells in the present case). This leads to the possibility of taking a spin structure from $Sk = 0$ to $Sk = 1$ by going through an energy barrier more linked to the micromagnetic energy balance than to topological considerations \cite{thiaville_rohart2013,PhysRevB.93.214412,stability_wiesendanger} which, in practice, allows intermediate values of $Sk$ to be found.

\begin{figure}
\centering
\includegraphics[width=0.75\textwidth]{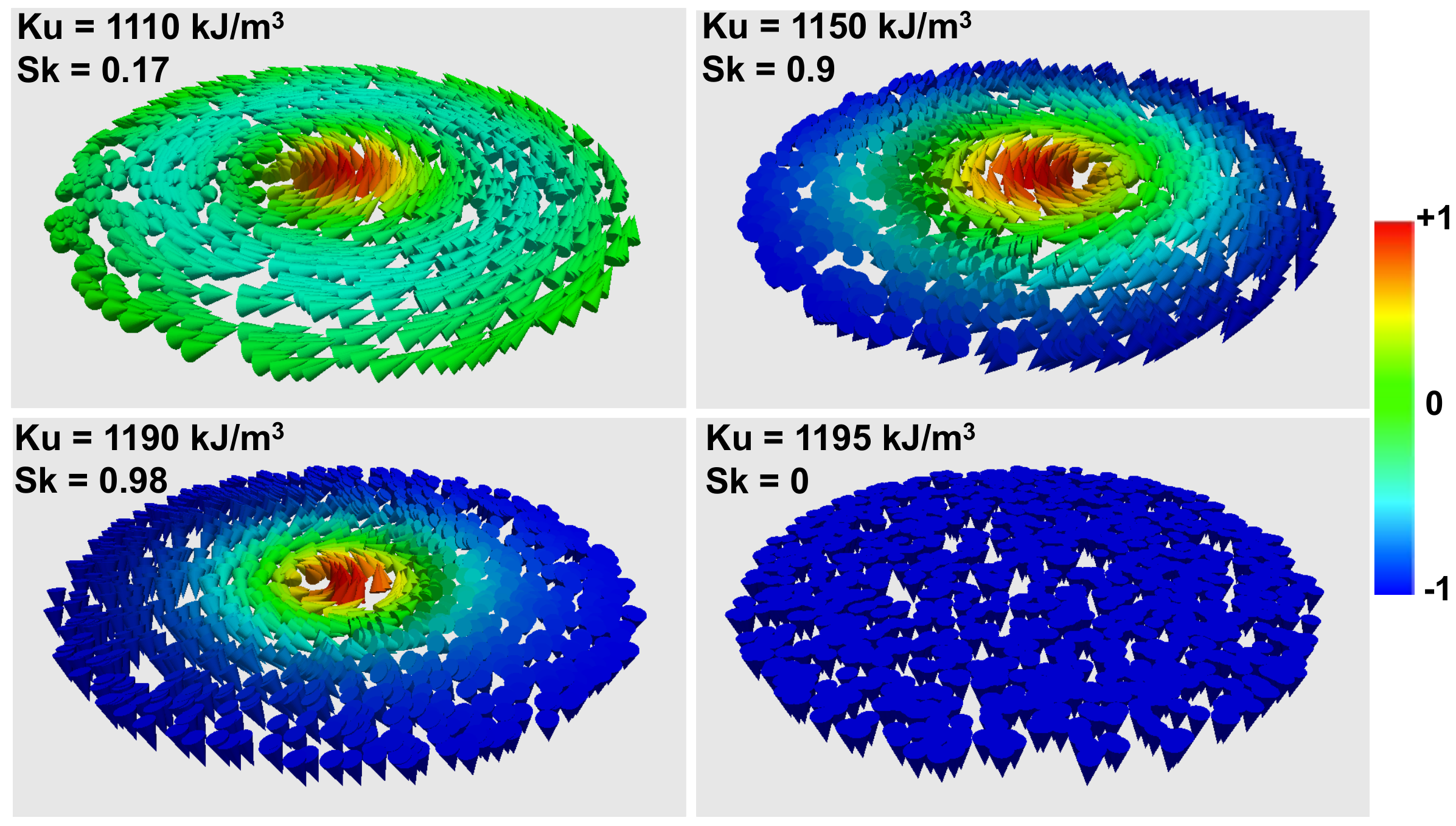}
\caption{Snapshots of magnetization for a given equilibrium state of a 128 nm cobalt disk with $D_{ex} = 0$ mJ/m$^2$. The $K_{u}$ values and the calculated skyrmion numbers are shown. The color bar gives the values of $m_z$.\label{fig:circle_snapshots}}
\end{figure}

\begin{figure}
\centering
\includegraphics[width=0.75\textwidth]{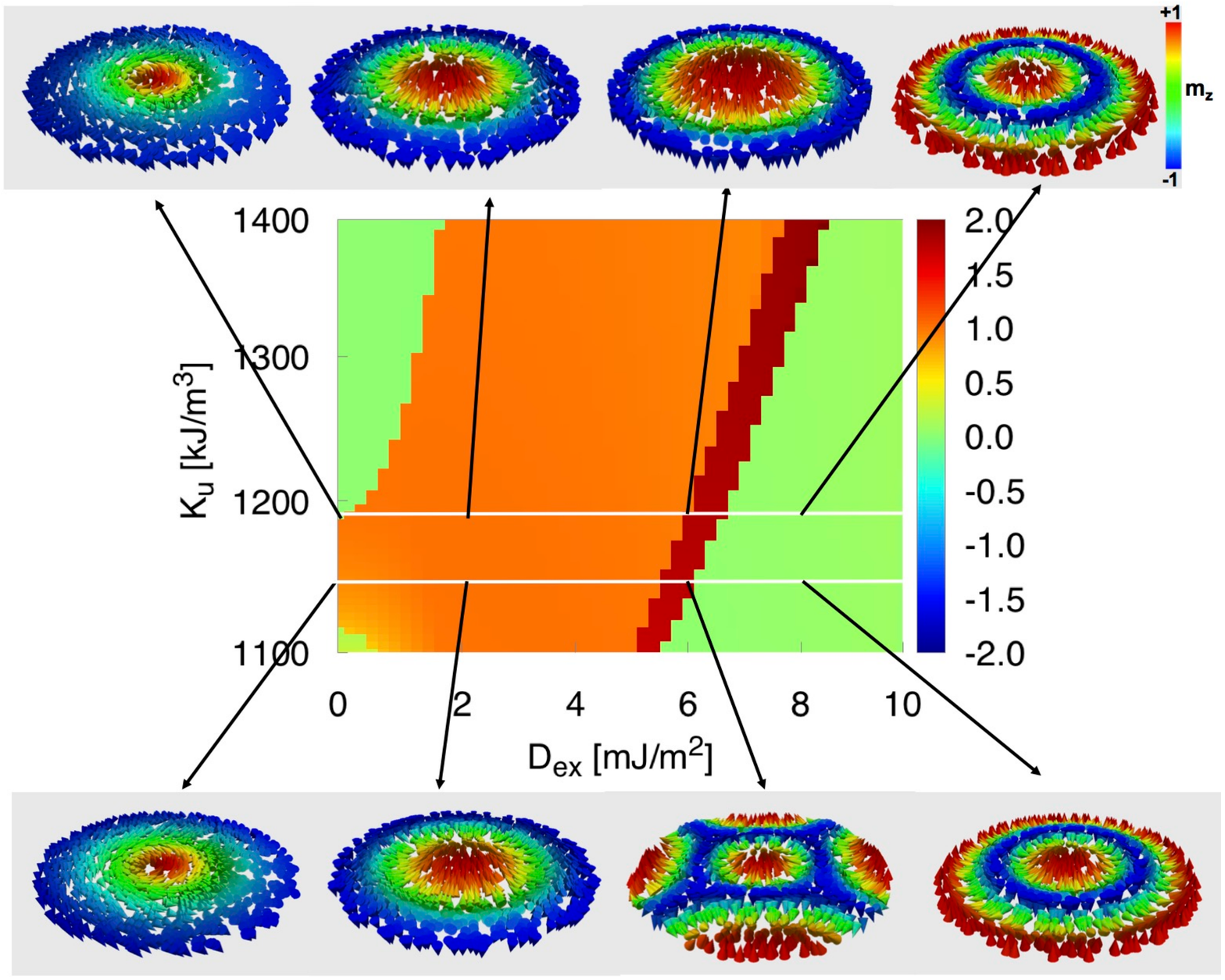}
\caption{Evolution of the magnetic structure of a $128$ nm disk for two values of uniaxial anisotropy constant ($K_{u} = 1150$ and $1180$ kJ/m$^3$) and several values of DMI (0, 2, 6 and 8 mJ/m$^2$). The initial state of the simulation was VL. As the DMI is increased, the disk evolves towards a HG skyrmion and multidomain state. A phase diagram from Fig. \ref{fig:maps_blochneel} is also shown. \label{fig:diag_cr}}
\end{figure}

\begin{figure}[h!]
\centering
\includegraphics[width=0.8\linewidth]{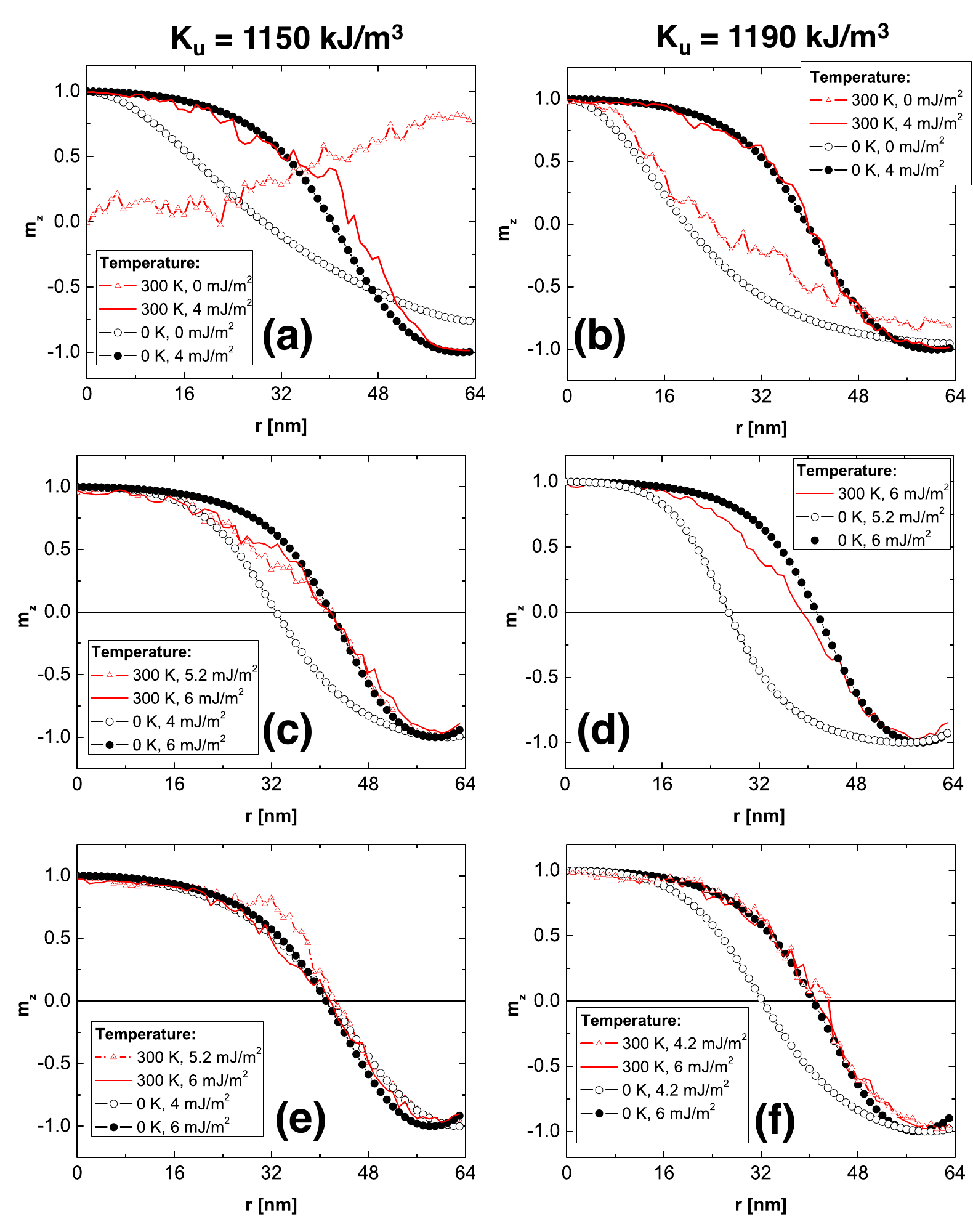}
\caption{Skyrmion profiles ($m_z \, \mathrm{vs.}\, r$) in a $128$ nm disk for different values of $D_{ex}$, $K_{u}$ and $T$. Profiles corresponding to each initial condition (VL, HG and uniform) are shown at $T = 0$ K (circles, empty or full) and $T = 300$ K (lines). All $300$ K profiles were averaged over several acquisitions. (a), (b): VL initial state, $K_{u} = 1150$ and $1190$ kJ/m$^3$, respectively; (c), (d): HG initial state, $K_{u} = 1150$ and $1190$ kJ/m$^3$, respectively; (e) and (f): Uniform initial state, $K_{u} = 1150$ and $1190$ kJ/m$^3$, respectively. Profiles of skyrmions in 64 nm disks are similar and are not shown.}
\label{fig:profiles}
\end{figure}

The diagrams show that, as a general rule, the presence of DMI (a finite value of $D_{ex}$) is a necessary condition for the observation of skyrmion ground states in these disks, while the requirements on anisotropy strength seem to be less stringent. However, diagrams of 128 nm disks initially in a VL state (Fig. \ref{fig:maps_blochneel}(g)) display a small $Sk = 1$ region, within a narrow range of $K_{u}$ values, for which $D_{ex} = 0$ mJ/m$^2$. This indicates that a stable skyrmion ground state can be reached in the absence of net DMI. Snapshots of a 128 nm disk along the transition to a zero-DMI skyrmion state as $K_{u}$ is increased are shown in Fig. \ref{fig:circle_snapshots}. The disk goes from a ground state with $Sk < 1$ to a VL skyrmion, with $Sk = 1$, as $K_{u}$ approaches $1990$ kJ/m$^3$. For higher $K_{u}$, the disk abruptly switches to a fully saturated uniform state, in line with the diagram in Fig. \ref{fig:maps_blochneel}(g). It is interesting that even though interfacial DMI, as applied in the simulations, is known to induce HG skyrmions in thin films \cite{sampaiocros2013}, the zero-DMI skyrmions observed are VL, typically induced by bulk DMI\cite{sampaiocros2013,sun2013}, or magnetostatic interactions\cite{review_finocchio}, which should be the present case. Increasing the value of $D_{ex}$ causes the VL structure to evolve towards HG, as expected from this interfacial DMI in thin films (Fig. \ref{fig:diag_cr} shows this transition, as the DMI is increased, for two values of $K_{u}$). This transition is also observed in $128$ nm squares (data not shown), and suggests that zero-DMI skyrmions appear as a result of the interplay between magnetostatic, exchange and anisotropy energies in these systems. It will be shown that this state is significantly less robust than skyrmions arising when the system has finite DMI, being more easily destroyed by magnetic fields or electric currents. Interestingly, this state persists at room temperature (Figs. \ref{fig:maps_128_temp} and \ref{fig:profiles}(b)). Zero-DMI skyrmions have been previously reported in the literature\cite{garcia2010,sun2013,PhysRevB.88.054403,xie_zerodmi}. In ref. \cite{xie_zerodmi}, the skyrmion ground state resulted from a complex interplay of anisotropy, exchange and magnetostatic interactions specific to the Co/Ru/Co multilayers studied, while in \cite{garcia2010}, the authors were able to observe, using photoelectron emission microscopy (PEEM), a HG skyrmion in a Co/Pt disk at room temperature within a narrow range of anisotropy values. In\cite{sun2013,PhysRevB.88.054403}, magnetostatic coupling to overlying nanostructures is exploited to stabilized the skyrmion.

%%%%%%%%%%%%%%%%%% Figure 2 %%%%%%%%%%%%%%%
%%%%%%%%%%%%%%%%%%%%%%%%%%%%%%%%%%%%%

%%%%%%%%%%%%%%%%%% Figure 3 %%%%%%%%%%%%%%
%%%%%%%%%%%%%%%%%%%%%%%%%%%%%%%%%%%%%

\begin{figure}
\centering
\includegraphics[width=0.8\textwidth]{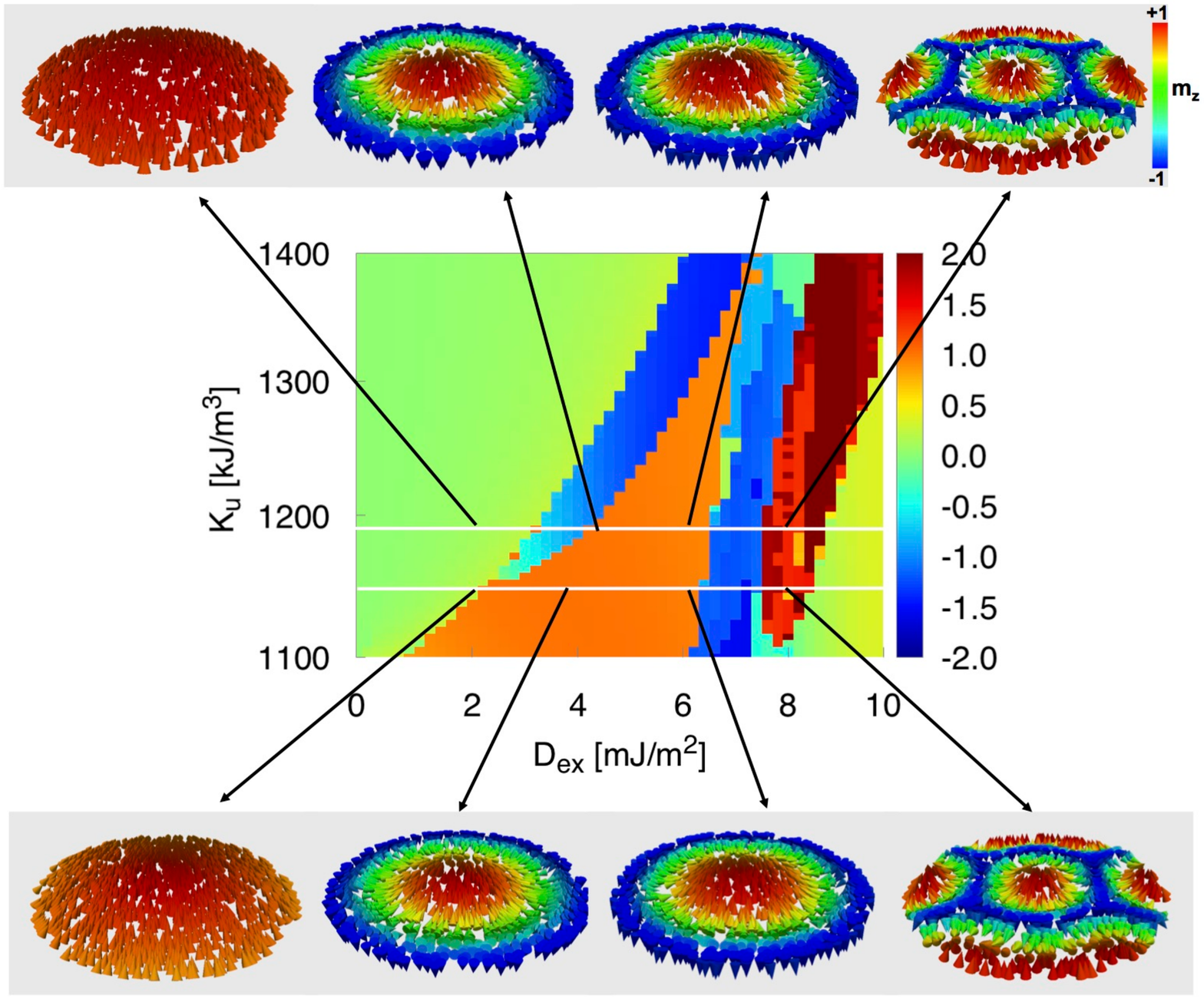}
\caption{Evolution of the magnetic structure of a $128$ nm disk for two values of uniaxial anisotropy constant ($K_{u} = 1150$ and $1180$ kJ/m$^3$) and several values of DMI (0, 2, 4 and 6 mJ/m$^2$). The disk was initially in a uniform state with $m_{z} = +1$, and as the DMI is increased, the disk evolves towards a HG skyrmion and multidomain state.  A phase diagram from Fig. \ref{fig:maps_blochneel} is also shown. \label{fig:diag_cr_saturated}}
\end{figure}

The simulations starting with the disk in a uniform state (Fig. \ref{fig:maps_blochneel}, (c), (f) and (i)) yield phase diagrams resembling the phase diagrams (e) and (h) (HG initial state). In the absence of DMI, all the disks initially uniform stay uniform due to the strong uniaxial out-of-plane anisotropy present. But as $D_{ex}$ is increased, a more complex diagram emerges, with stable skyrmions, multidomain and uniform states appearing for certain combinations of $D_{ex}$ and $K_{u}$. The similarity to HG diagrams can be understood from a more careful analysis of the time evolution of the disk magnetization from a initial HG skyrmion at $t=0$ s. Once the simulation starts, this skyrmion quickly collapses, leaving the disk in a uniform state which evolves towards the final equilibrium states in the diagram. This shows that the time evolution of the simulations with HG or uniform initial states are practically the same and lead to very similar diagrams of Figs. \ref{fig:maps_blochneel}(e) and (f) or (h) and (i). The slight differences are ascribed to the different pathways in parameter space followed by the system until it reaches its ground state (and to a few cases in which the initial skyrmion might not have collapsed). In particular, the transition from the uniform to the skyrmion state involves overcoming the energy barrier associated with changing $Sk$ from $0$ to $1$. Snapshots of the magnetic structures of $128$ nm disks initially uniform, corresponding to several combinations of $K_{u}$ and $D_{ex}$, show that the skyrmions  appearing for $D_{ex} > 0$ are always HG (Figure \ref{fig:diag_cr_saturated}), as reported in the literature. If the initial state is VL, the skyrmion generally does not collapse into a uniform state, but evolves towards HG, favored by the interfacial DMI term ($D_{ex}$) of the simulations.

%%%%%%%%%%%%%%%% Figure 4 %%%%%%%%%%%%%%%
%%%%%%%%%%%%%%%%%%%%%%%%%%%%%%%%%%%%%

\subsection{Skyrmion stability with finite temperature} \label{sec:temp}

\begin{figure}[h!]
\centering
\includegraphics[width=0.8\linewidth]{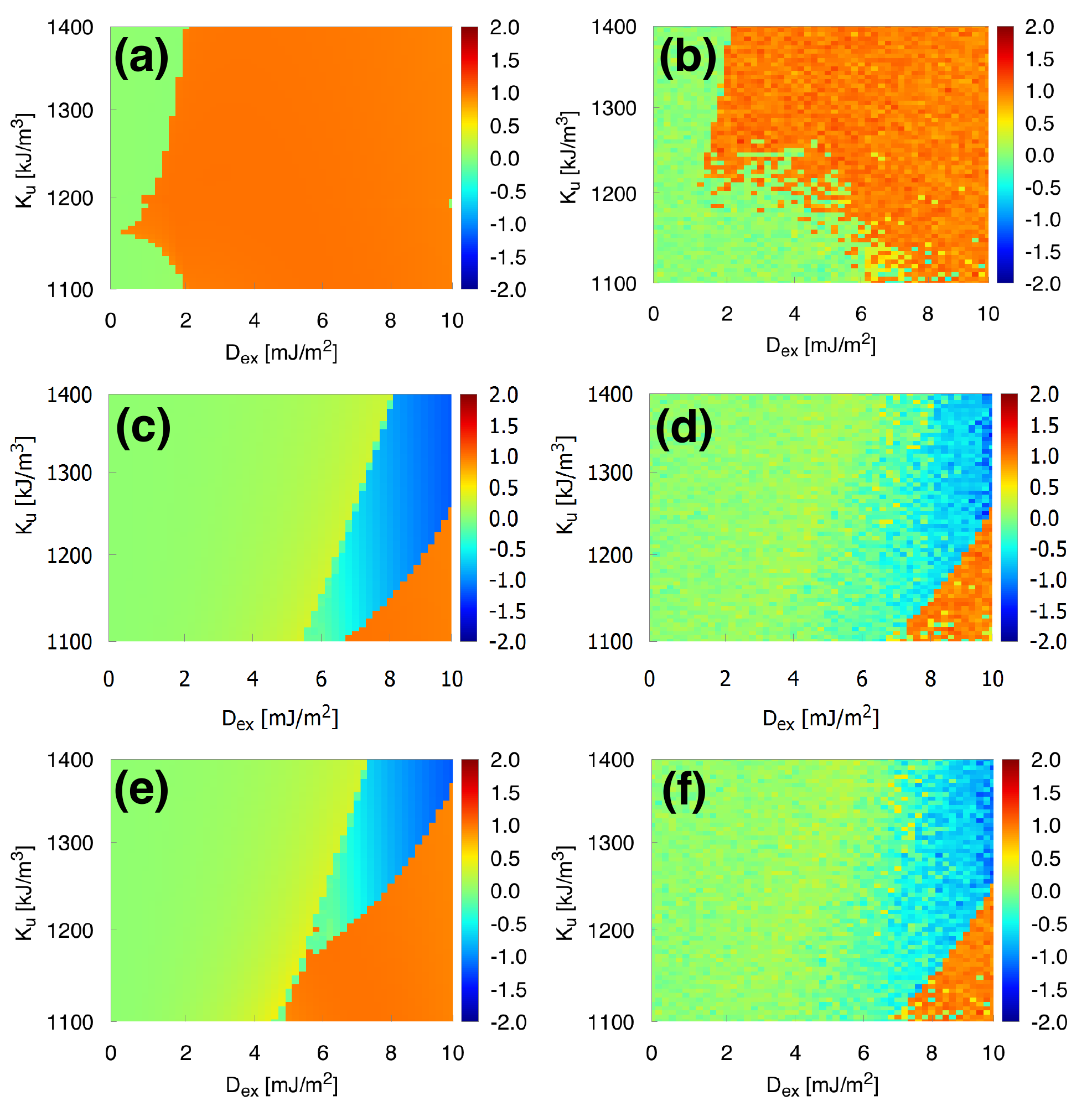}
\caption{Skyrmion phase diagrams of $64$ nm disks with a VL ((a), (b), first row), HG ((c), (d), second row) and uniform state ((e), (f), third row) as initial configurations, simulated with $T = 0$ K (left column) and $T = 300$ K (right column). Each diagram represents a region of the parameter space where $K_{u}$ is comprised in the $1100$ -- $1400$ kJ/m$^3$ range and $D_{ex}$ is comprised between $0$ and $10$ mJ/m$^2$. In each diagram, the color indicates the skyrmion number \emph{Sk}, with orange representing the stable skyrmion ground state ($Sk = 1$). (Color online.)} 
\label{fig:maps_64_temp}
\end{figure}

\begin{figure}[h!]
\centering
\includegraphics[width=0.8\linewidth]{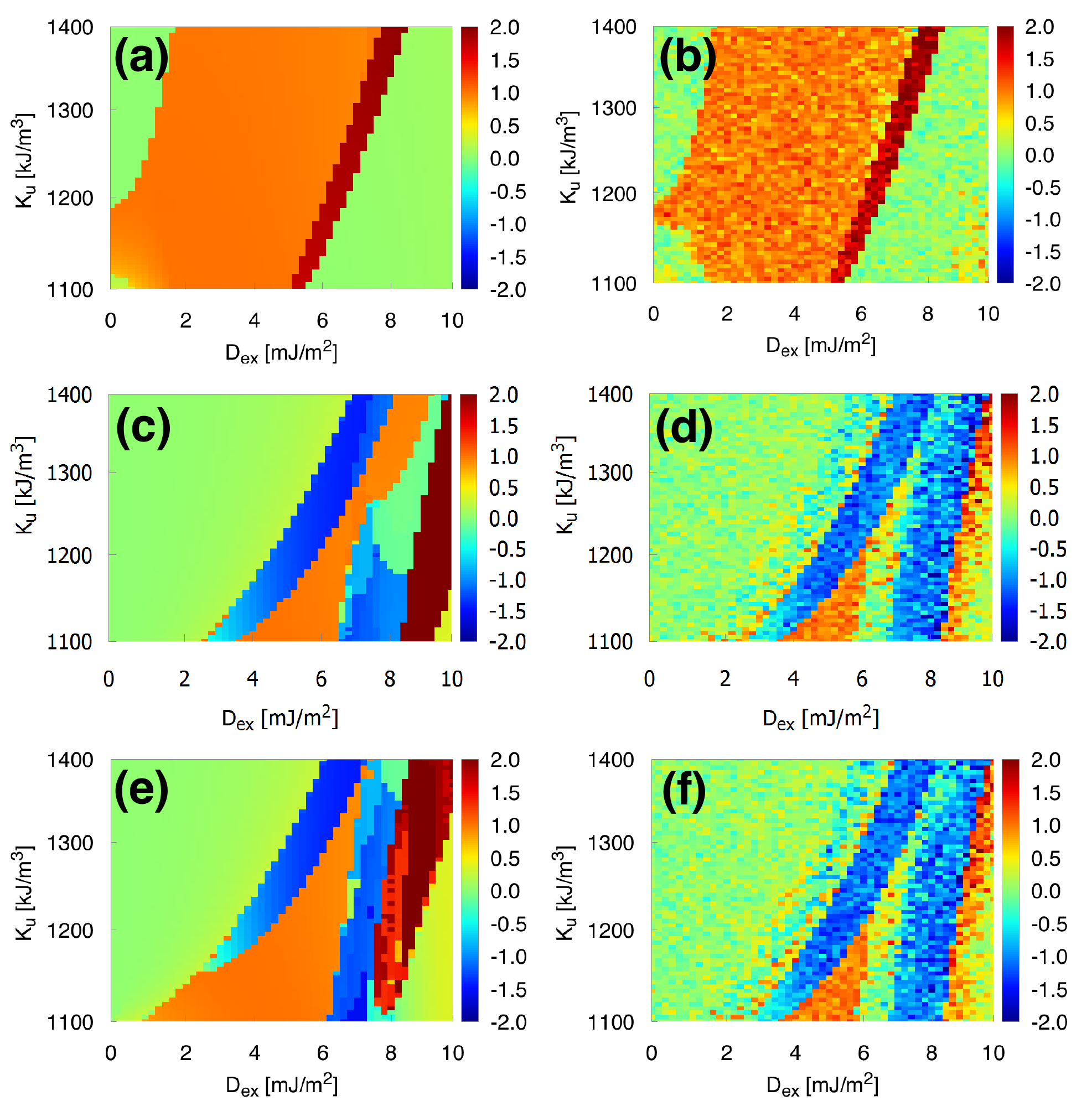}
\caption{Skyrmion phase diagrams of $128$ nm disks with a VL ((a), (b), first row), HG ((c), (d), second row) and uniform state ((e), (f), third row) as initial configurations, simulated with $T = 0$ K (left column) and $T = 300$ K (right column). Each diagram represents a region of the parameter space where $K_{u}$ is comprised in the $1100$ -- $1400$ kJ/m$^3$ range and $D_{ex}$ is comprised between $0$ and $10$ mJ/m$^2$. In each diagram, the color indicates the skyrmion number \emph{Sk}, with orange representing the stable skyrmion ground state ($Sk = 1$). (Color online.)}
\label{fig:maps_128_temp}
\end{figure}

A finite temperature of $300$ K was introduced in the simulations in order to study the robustness of the skyrmion ground states at room temperature. Thus, the phase diagrams obtained are more fit for comparison with experimental results. The $300$ K phase diagrams of $64$ and $128$ nm disks are shown in Figs. \ref{fig:maps_64_temp} and \ref{fig:maps_128_temp} along with the respective $0$ K diagrams for comparison (diagrams of $32$ nm disks are not shown because no skyrmion ground state was observed at $300$ K in these disks). The finite temperature has two effects on the skyrmion ground states: \emph{(i)} regions where skyrmions are stable at $0$ K are smaller at $300$ K, sometimes even disappearing completely; \emph{(ii)} fuzzier borders now separate different regions in the diagrams, suggesting that the phases (skyrmionic, uniform or multidomain) are no longer well defined for the combinations of $K_u$ and $D_{ex}$ along these borders. As an example, the stability range of a $128$ nm disk with a uniform initial state and $K_{u} = 1190$ kJ/m$^3$ goes from $4.2 \textrm{---} 6.4$ mJ/m$^2$ at $0$ K to $5.2 \textrm{---} 6.0$ mJ/m$^2$ at $300$ K, a general effect observed for all combinations of $D_{ex}$ and $K_{u}$ investigated. Thermally-induced random switching between skyrmionic and uniform phases has been observed in Monte Carlo simulations of hexagonal ferromagnetic structures mimicking Pd/Fe/Ir(111) samples at temperatures close to a critical value\cite{stability_wiesendanger}. The switching between phases occurred when the thermal energy was comparable to the micromagnetic energy barrier separating the skyrmion from the uniform phase. In that work, the value of $D_{ex}$ was kept constant while the external magnetic field and temperature were varied. Nonetheless, the random switching between phases induced by finite temperature is in line with our observation of less well defined phase boundaries separating the skyrmionic phase from uniform or multidomain states. The overall prevalence of the skyrmionic phase observed in our results is also in line with these results\cite{stability_wiesendanger}, which revealed a much smaller attempt frequency for skyrmion annihilation than for skyrmion creation, when the thermal dependence of the skyrmion lifetimes are analyzed by means of an Arrhenius law. This supports our finding that, despite the smaller $Sk = 1$ areas in the $T = 300$ K phase diagrams, they do not disappear completely, showing a good potential for skyrmion stability at room temperatures for these combinations of $K_{u}$ and $D_{ex}$. It is important to highlight that in \cite{stability_wiesendanger} the authors conclude by proposing the use of multilayers in order to increase the exchange coupling ($J$) and the DMI. Increasing these parameters would make stable room temperature skyrmions viable, according to their findings, and in this work we are exploring skyrmion stability for material parameter values typically found in Co thin film-based systems.

%%%%%%%%%%%%%%%% Figure 5 %%%%%%%%%%%%%%%
%%%%%%%%%%%%%%%%%%%%%%%%%%%%%%%%%%%%

%%%%%%%%%%%%%%%% Figure 6 %%%%%%%%%%%%%%%
%%%%%%%%%%%%%%%%%%%%%%%%%%%%%%%%%%%%

%%%%%%%%%%%%%%%% Figure 7 %%%%%%%%%%%%%%%
%%%%%%%%%%%%%%%%%%%%%%%%%%%%%%%%%%%%

\subsection{Skyrmion sizes} \label{sec:size}

\begin{figure}
\centering
\includegraphics[width=0.8\textwidth]{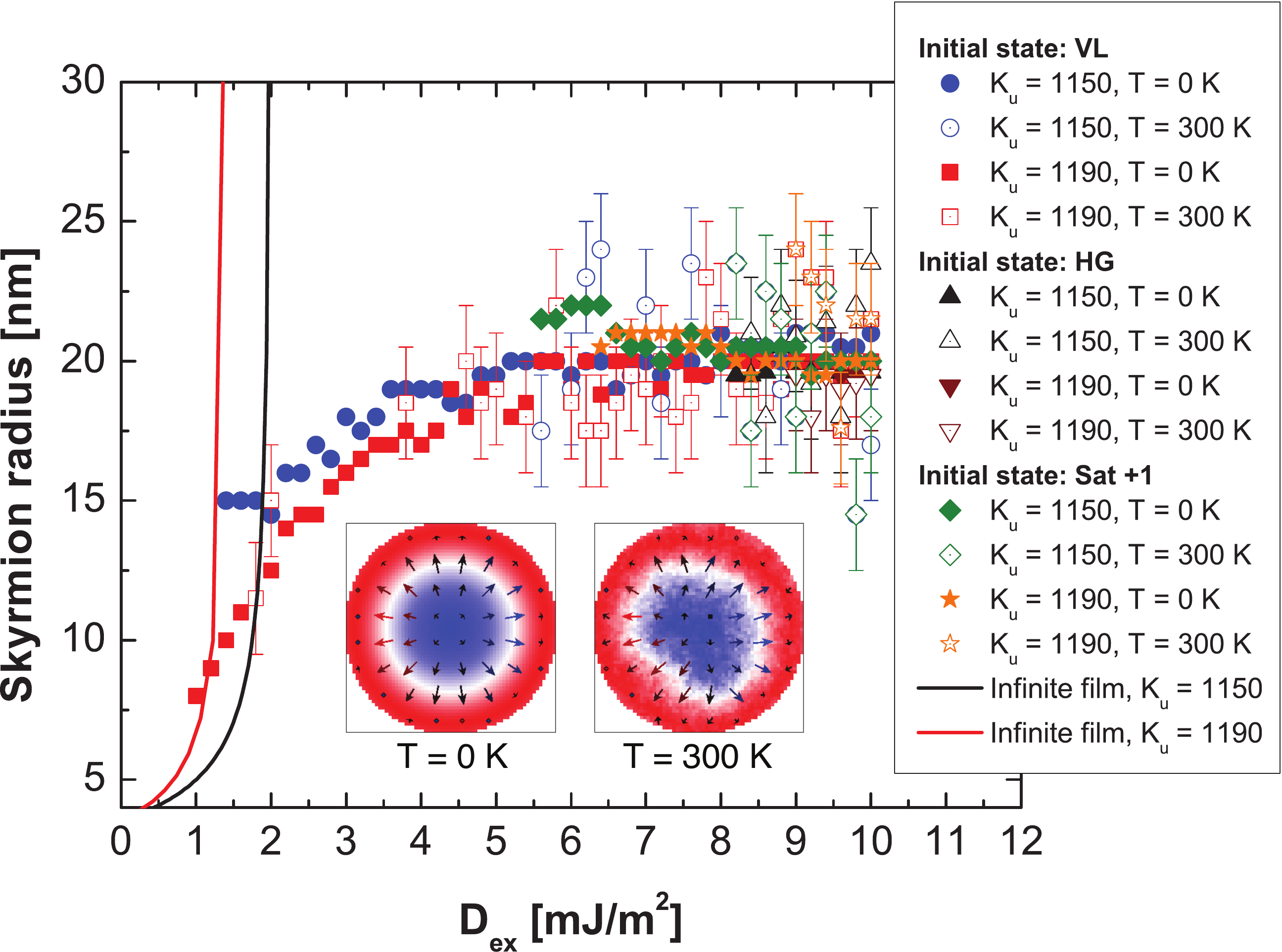}
\caption{Evolution of the skyrmion radius in 64 nm disks for all initial conditions studied, $K_{u} = 1150$ and $1190$ kJ/m$^3$, $T = 0$ and $300$ K. The lines show the behavior of skyrmions in infinite films, for which a $0$ K analytical solution to the micromagnetic problem is possible (Eq. (\ref{eq:radius_inf})). In the insets, a skyrmion in a disk with $K_u = 1150$ kJ/m$^3$ and $D_{ex} = 8.8$ mJ/m$^2$ is shown for $T = 0$ K (left) and $T = 300$ K (right). ($m_z = +1 \rightarrow$ red; $m_z = -1 \rightarrow$ blue).  \label{fig:radius_dex_64}}
\end{figure}

\begin{figure}
\centering
\includegraphics[width=0.8\textwidth]{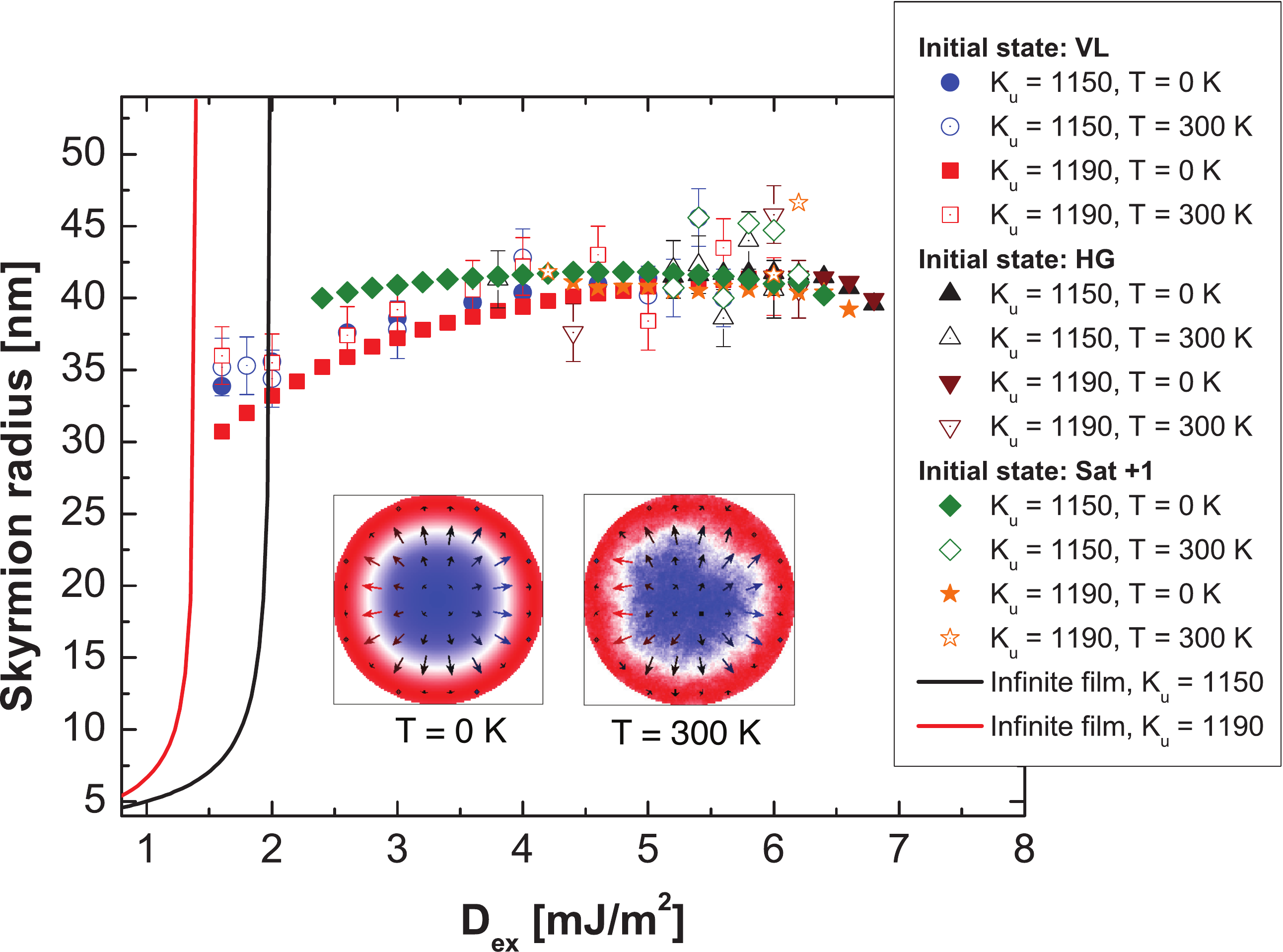}
\caption{Evolution of the skyrmion radius in 128 nm disks for all initial conditions studied, $K_{u} = 1150$ and $1190$ kJ/m$^3$, $T = 0$ and $300$ K. The lines show the behavior of skyrmions in infinite films, for which a $0$ K analytical solution to the micromagnetic problem is possible (Eq. (\ref{eq:radius_inf})). In the insets, a skyrmion in a disk with $K_u = 1190$ kJ/m$^3$ and $D_{ex} = 6.2$ mJ/m$^2$ is shown for $T = 0$ K (left) and $T = 300$ K (right). ($m_z = +1 \rightarrow$ red; $m_z = -1 \rightarrow$ blue). \label{fig:radius_dex}}
\end{figure}

The average skyrmion core radius was determined from skyrmion profiles in order to analyze the evolution of skyrmion size with $D_{ex}$ at both temperatures for $64$ and $128$ disks. The profiles are plots of reduced magnetization $m_{z}$ along the disk radius (Figs. \ref{fig:profiles}) for disks with $K_{u} = 1150$ and $1190$ kJ/m$^3$. The radius is defined as the distance from the disk center to the $m_z = 0$ line. If $T = 300$ K, several profiles along different radial cuts are used to determine the average radius. The evolution of the core radius within the corresponding stability regions is plotted against $D_{ex}$ for several initial states and both temperatures in Figs. \ref{fig:radius_dex_64} (64 nm) and \ref{fig:radius_dex} (128 nm). The lines show the expected behavior for an infinite film, where an analytical solution of the micromagnetic problem is possible, and yields the following expression for the skyrmion core radius \cite{thiaville_rohart2013}:

%%%%%%%%%%%%%%%%%%%%%%%%%
\begin{equation}
R_{s} \approx \frac{\Delta}{\sqrt{2 (1- D/D_c)}}
\label{eq:radius_inf}
\end{equation}
%%%%%%%%%%%%%%%%%%%%%%%%%
where $D_c = 4 \sqrt{A K} / \pi$ is the critical DMI energy constant above which domain walls will appear in an infinite film \cite{thiaville_rohart2013} and $\Delta = \sqrt{A/K}$ is the theoretical domain wall width parameter.

The results show that the core radius tends to fall around the same average value, corresponding to $60 \%$ of the disk radius, regardless of disk size, a result of the confinement effect imposed by the disk finite diameter \cite{thiaville_rohart2013,sampaiocros2013}. Furthermore, the radius evolution with $D_{ex}$ is the same for all initial conditions and temperatures, indicating the dominance of the DMI in the formation and stability of these spin structures. Notably, within the stability ranges, temperature does not alter significantly the radius of skyrmions in the disks. As expected, the skyrmion radius increases with DMI strength \cite{thiaville_rohart2013}, but in this case it saturates due to the confinement effect. Figs. \ref{fig:radius_dex_64} and \ref{fig:radius_dex} also show the edge effect on the skyrmion radius as a slight decrease in the skyrmion core radius before its disappearance for high DMI values. Due to confinement inside the disk, the skyrmion radius is limited by the edges and decreases slightly before the skyrmion transitions to more complex non-collinear structures favored by the higher DMI \cite{beg_scirep,thiaville_rohart2013,sampaiocros2013}. Finally, the finite temperature partially suppresses, but does not eliminate, the zero-DMI skyrmion ground states of 128 nm disks (Figs. \ref{fig:maps_128_temp}(b) and \ref{fig:profiles}(b)).

%%%%%%%%%%%%%%%% Figure 8 %%%%%%%%%%%%%%
%%%%%%%%%%%%%%%%%%%%%%%%%%%%%%%%%%%

%%%%%%%%%%%%%%%% Figure 9 %%%%%%%%%%%%%%%
%%%%%%%%%%%%%%%%%%%%%%%%%%%%%%%%%%%

\subsection{Skyrmion stability under applied magnetic fields} \label{sec:field}

The stability of the skyrmion ground states achieved in 64 and 128 nm disks was tested against out-of-plane magnetic fields by applying positive and negative (parallel/antiparallel to skyrmion core magnetization) magnetic field on the disks. The field magnitude is increased in 10 Oe steps from $0$ Oe to the critical value ($H_{crit,+-}$) where skyrmion annihilation is observed. These tests were done both at $T = 0$ K and $T = 300$ K. At $T = 0$ K, a conjugate gradient energy minimization routine is used at each field step, but for $T = 300$ K this routine is not available in mumax3, so each field step must be applied during a fixed time ($\gtrsim 5$ ns).

%VL (with $D_{ex} = 0$ mJ/m$^2$) and HG skyrmions, with $K_{u} = 1190$ kJ/m$^3$ and different values of $D_{ex}$ ($2, 3, 4$ and $5$ mJ/m$^2$), 

\begin{figure}
\centering
\includegraphics[width=.8\textwidth]{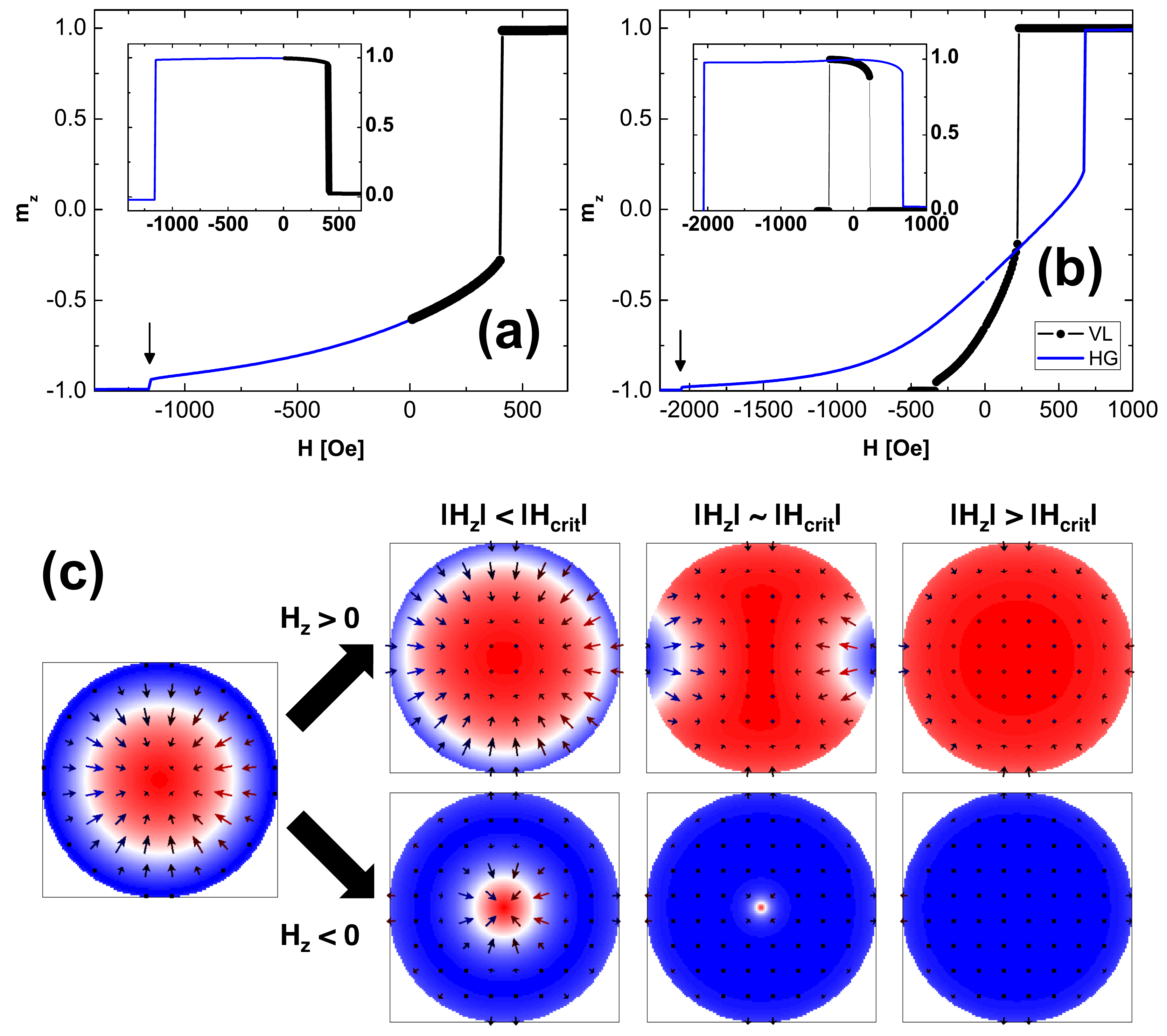}
\caption{Field-driven evolution of the out-of-plane component of the reduced magnetization, $m_{z}$, of: (a) a 64 nm disk from a HG initial state with $D_{ex} = 2$ mJ/m$^2$; (b) a 128 nm disk from two initial states: VL, with $D_{ex} = 0$ mJ/m$^2$, and HG with $D_{ex} = 2$ mJ/m$^2$. $K_{u} = 1190$ MJ/m$^3$ in all cases. The insets shows the evolution of the skyrmion number. (c) Snapshots of the disk magnetization evolving towards annihilation for positive and negative fields ($m_z = +1 \rightarrow$ red; $m_z = -1 \rightarrow$ blue). (Color online.) \label{fig:ann_fields_pos}}
\end{figure}

For fields parallel to the skyrmion core (positive fields), the diameter of the skyrmion core always increases, until a critical value ($H_{crit,+}$) where the skyrmion core touches an edge of the disk and forms a highly unstable domain wall that quickly disappears (Fig. \ref{fig:ann_fields_pos}). Interestingly, for higher values of DMI, the skyrmion would in some cases evolve towards a 360$^{\circ}$ domain wall (effectively a pair of 180$^{\circ}$ N\'eel walls with same chirality). The disk would then remain in this highly stable configuration at a fixed value of field, only reaching the uniform state for much larger fields. When negative fields are applied, the skyrmion core shrinks continuously until its complete disappearance at $H_{crit,-}$, as shown in Fig. \ref{fig:ann_fields_pos}. Noticeably, the magnitude of the negative critical fields for skyrmion annihilation are substantially larger in a few cases: $|H_{crit,-}| > |H_{crit,+}|$, which is explained by the magnetostatic repulsion of the edges of the skyrmion core as it shrinks. It is important to stress that no modifications in skyrmion structure (VL or HG) have been observed under applied magnetic field, and that no skyrmion motion has been observed during the process. The absolute value of the annihilation field is always larger for HG skyrmions, with $D_{ex} > 0$ mJ/m$^2$, indicating that the energy barrier for skyrmion annihilation is larger in this case, in line with results in the literature showing HG as the stable skyrmion topology predominantly found in magnetic ultrathin films \cite{review_finocchio,sampaiocros2013}. While VL skyrmions can appear with $D_{ex} = 0$ mJ/m$^2$ within narrow $K_{u}$ ranges, they are much less stable than HG skyrmions, being easily destroyed by out-of-plane fields as low as $230$ Oe, while the lowest annihilation field of an HG skyrmion is $680$ Oe.

\begin{figure}
\centering
\includegraphics[width=\textwidth]{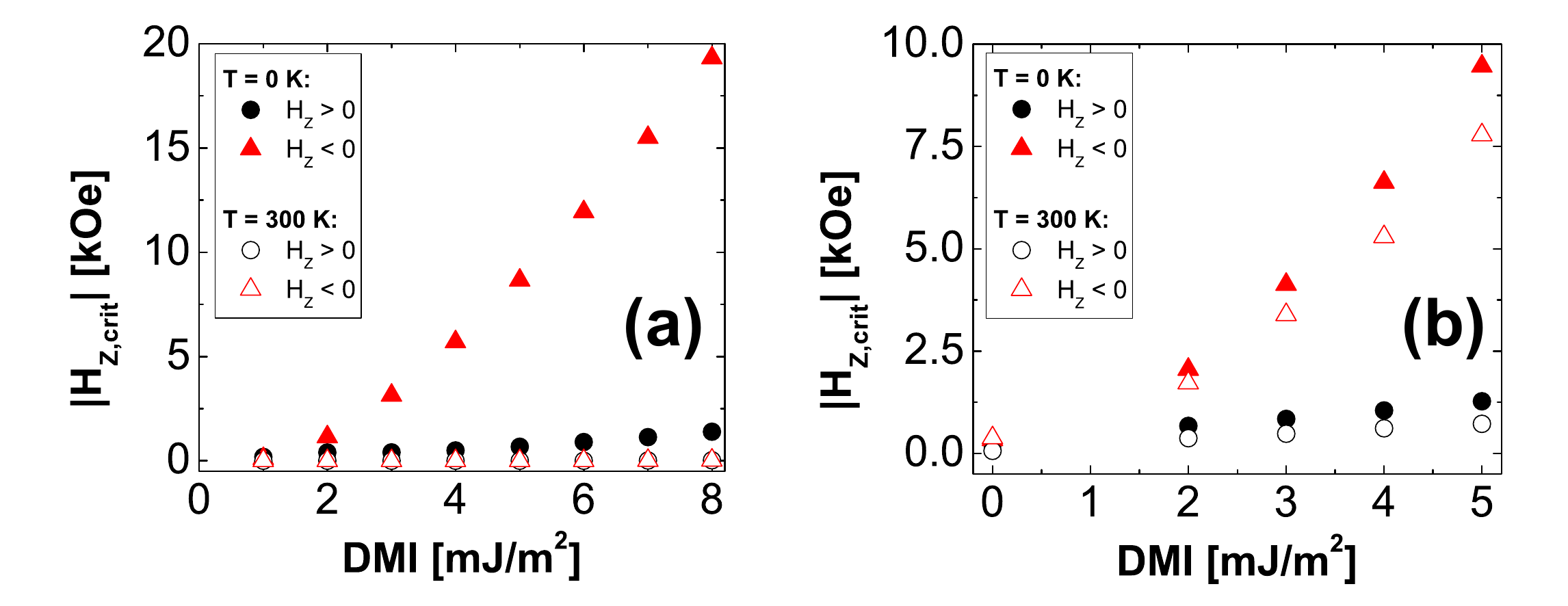}
\caption{Critical out-of-plane magnetic fields for skyrmion annihilation as a function of DMI strength determined for (a) 64 and (b) 128 nm disks at $T = 0$ K (full symbols) and $T = 300$ K (open symbols). (Color online.)\label{fig:crit_field_dmi}}
\end{figure}

At room temperature ($T = 300$ K), the same overall behavior is observed: positive fields lead to skyrmion annihilation after its core reaches the disk edge and negative fields shrink the core until it disappears, the former process occurring at lower field magnitudes than the latter. When compared to 0 K, all the critical field magnitudes at $300$ K are lower. The critical fields are shown in Table \ref{tab:crit_fields}, and their dependence with DMI in shown in Fig. \ref{fig:crit_field_dmi}. For both disk sizes, the annihilation fields are lower at 300 K, but this reduction is drastic (two orders of magnitude) for 64 nm disks, evidencing a much lower energy barrier separating the skyrmion state from the uniform state at this size. Another interesting fact observed in 64 nm disks is that the asymmetry between positive and negative critical fields vanishes at $300$ K .

%%%%%%%%%%%%%%%% Figure 10 %%%%%%%%%%%%%%%
%%%%%%%%%%%%%%%%%%%%%%%%%%%%%%%%%%%%

%%%%%%%%%%%% Table 1 - Critical fields %%%%%%%%%%%%
\begin{table}[h!]
\centering
\begin{tabular}{| C{3cm} | C{3cm} | C{3cm} | C{3cm} | C{3cm} |}
%\hline
  %& \multicolumn{3}{|c|}{Primeiro ms} & \multicolumn{3}{|c|}{Segundo ms} \\
 \hline
  & \multicolumn{2}{|c|}{\textbf{$\mathbf{d = }$ 64 nm}} & \multicolumn{2}{|c|}{\textbf{$\mathbf{d = }$ 128 nm}} \\
 \hline
 $\mathbf{D_{ex}}$ \textbf{(mJ/m$^2$)} & $\mathbf{H_{crit,+}^{0 \mathrm{K}}}/ \mathbf{H_{crit,+}^{300 \mathrm{K}}}$ & $\mathbf{H_{crit,-}^{0 \mathrm{K}}}/ \mathbf{H_{crit,-}^{300 \mathrm{K}}}$ & $\mathbf{H_{crit,+}^{0 \mathrm{K}}}/ \mathbf{H_{crit,+}^{300 \mathrm{K}}}$ & $\mathbf{H_{crit,-}^{0 \mathrm{K}}}/ \mathbf{H_{crit,-}^{300 \mathrm{K}}}$ \\
\hline
\hline
$\mathbf{0} \,\mathrm{(VL)}$ & $--$ & $--$ & $220/60$ & $-330/-380$ \\
\hline
$\mathbf{1.0}$ & $180/0$ & $-110/0$ & $--$ & $--$ \\
\hline
%\multirow{4}{*}{\textbf{HG}} &
$\mathbf{2.0}$ & $400/1$ & $-1150/0$ & $670/450$ & $-2050/-1720$ \\
%\cline{2-4}
\hline
$\mathbf{3.0}$ & $410/1$ & $-3140/0$ & $840/500$ & $-4130/-3390$ \\
%\cline{2-4}
\hline
$\mathbf{4.0}$ & $500/1$ & $-5700/-1$ & $1050/700$ & $-6620/-5290$ \\
%\cline{2-4}
\hline
$\mathbf{5.0}$ & $680/1$ & $-8660/-1$ & $1270/850$ & $-9460/-7780$ \\
\hline
$\mathbf{6.0}$ & $900/2$ & $-11940/-2$ & $--$ & $--$ \\
\hline
$\mathbf{7.0}$ & $1140/16$ & $-15500/-15$ & $--$ & $--$ \\
\hline
$\mathbf{8.0}$ & $1390/30$ & $-19320/-29$ & $--$ & $--$ \\
\hline
\end{tabular}
\caption{Critical magnetic fields for skyrmion annihilation in 64 nm and 128 nm disks, at $T = $ 0  and 300 K. All field values in kOe.}
\label{tab:crit_fields}
\end{table}
%%%%%%%%%%%%%%%%%%%%%%%%%%%%%%%%%%%%%%

%%%%%%%%%%%%%%%%%% Figure 11 %%%%%%%%%%%%%%%
%%%%%%%%%%%%%%%%%%%%%%%%%%%%%%%%%%%%%

\subsection{Skyrmion stability under spin polarized currents} \label{sec:current}

\begin{figure}
\centering
\includegraphics[width=.8\textwidth]{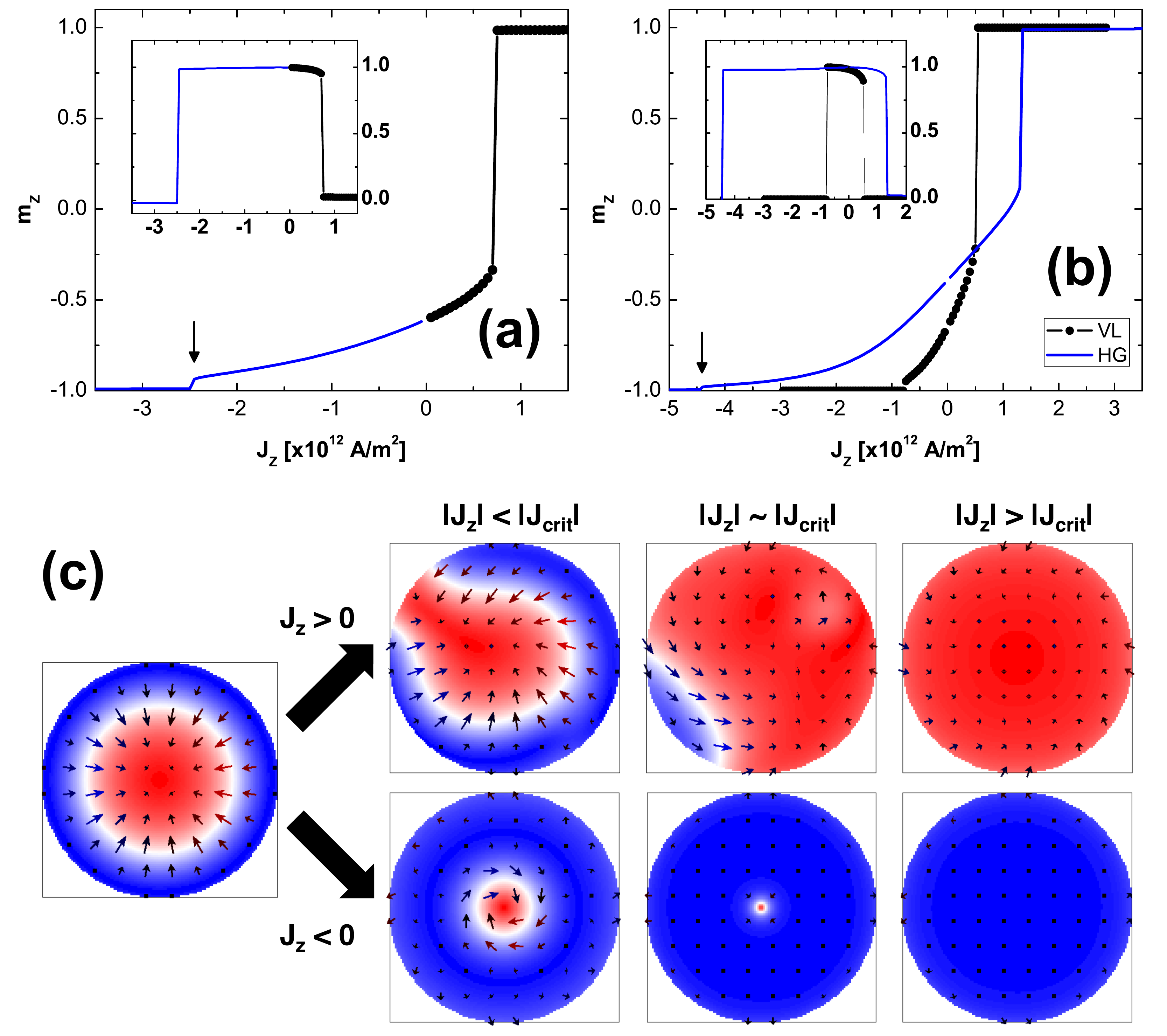}
\caption{Spin polarized current-driven evolution of the out-of-plane component of the reduced magnetization, $m_{z}$, of: (a) a 64 nm disk from a HG initial state with $D_{ex} = 2$ mJ/m$^2$; (b) a 128 nm disk from two initial skyrmion states: VL, with $D_{ex} = 0$ mJ/m$^2$, and HG with $D_{ex} = 2$ mJ/m$^2$. $K_{u} = 1190$ MJ/m$^3$ in all cases. The insets shows the evolution of the skyrmion number. All current values, unless otherwise noted, are multiples of $10^{12}$ A. (c) Snapshots of the disk magnetization evolving towards annihilation for positive and negative currents ($m_z = +1 \rightarrow$ red; $m_z = -1 \rightarrow$ blue). (Color online.) \label{fig:ann_curr_pos}}
\end{figure}

The effect of spin polarized electric currents, by means of Slonczewski-like STT, on the same structures tested in Secs. \ref{sec:temp}, \ref{sec:size} and \ref{sec:field} was investigated at $T = 0$ and $300$ K for the same values of $K_u$ and $D_{ex}$. Currents were applied along the z axis (perpendicular to the disk plane) with out-of-plane spin polarization of $0.4$ \cite{sampaiocros2013} and non-adiabaticity $\beta = 0.2$. At both temperatures, the initial value of the current density was always equal to $1$ x $10^{10}$ A/m$^2$ ($0.01$ TA/m$^2$, equivalent to $\approx 32$ and $\approx 129$ $\mu$A for 64 and 128 nm disks, respectively) and its magnitude was increased in $0.005$ TA/m$^2$ steps until the critical current density for skyrmion annihilation, $J_{crit,+-}$, was reached (Fig. \ref{fig:ann_curr_pos}). For 64 nm skyrmions at 300 K, the applied currents had to be drastically reduced (several orders of magnitude) due to the extreme instability of the skyrmions under electric current in these samples at room temperature.

For both disk diameters it was observed that, for low DMI, HG skyrmions always tend to a VL structure, regardless of the current sense (Figs. \ref{fig:ann_curr_pos}; similar transformations of skyrmion structure under applied currents have been previously reported \cite{PhysRevLett.114.137201,PhysRevB.93.024415}). At $T= 0$ K, positive currents (negative electron flow) cause the skyrmion core diameter to increase until annihilation, which occurs after the core touches the disk edge, forming an unstable domain wall, in analogy to the field-driven annihilation process previously observed. For negative currents, the skyrmion core diameter shrinks until its disappearance, which happens at critical currents with magnitudes much larger than the positive critical currents (again, in analogy to what was observed with applied magnetic fields). At $T = 300$ K, similar dynamics is observed, but the critical current magnitudes are always lower. Again, in the 64 nm disks, the reduction in critical current values is drastic at $300$ K, with critical currents 12 orders of magnitude below those observed in 128 nm disks under the same conditions. This amounts to very unstable skyrmions under current in these disks, especially if such current values are comparable to the noise level of the integrated electronic circuits where these disks are meant to operate. This shows that the size of the nanostructure supporting the skyrmion may have an enormous impact in its stability, with very small structures with $K_u$ and DMI values within the investigated ranges not feasible at room temperature. Finally, it is interesting to note that the skyrmion shrinking or expansion process under current (or magnetic field) is reversible, with the skyrmion returning to its original diameter once the current (or field) is removed. On the other hand, the skyrmion annihilation process is always irreversible, which means that if the current (or field) is removed after annihilation, the disk remains in a uniform state.

%($J_{crit}^{precession}$)
%%%%%%%%%%%%%%%%% Figure 12 %%%%%%%%%%%%%%%
%%%%%%%%%%%%%%%%%%%%%%%%%%%%%%%%%%%%%

%%%%%%%%%%%% Table 2 - Critical currents %%%%%%%%%%%%
\begin{table}[h!]
\centering
\begin{tabular}{| C{3cm} | C{3cm} | C{3cm} | C{3cm} | C{3cm} |}
%\hline
  %& \multicolumn{3}{|c|}{Primeiro ms} & \multicolumn{3}{|c|}{Segundo ms} \\
 \hline
  & \multicolumn{2}{|c|}{\textbf{64 nm}} & \multicolumn{2}{|c|}{\textbf{128 nm}} \\
 \hline
 $\mathbf{D_{ex}}$ \textbf{(mJ/m$^2$)} & $\mathbf{J_{crit,+}^{0 \mathrm{K}}}$/$\mathbf{J_{crit,+}^{300 \mathrm{K}}}$ & $\mathbf{J_{crit,-}^{0 \mathrm{K}}}$/$\mathbf{J_{crit,-}^{300 \mathrm{K}}}$ & $\mathbf{J_{crit,+}^{0 \mathrm{K}}}$/$\mathbf{J_{crit,+}^{300 \mathrm{K}}}$ & $\mathbf{J_{crit,-}^{0 \mathrm{K}}}$/$\mathbf{J_{crit,-}^{300 \mathrm{K}}}$ \\
\hline
\hline
$\mathbf{0} \,\mathrm{(VL)}$ & $--$ & $--$ & $0.5/0.56$ & $-0.75/-0.12$ \\
\hline
$\mathbf{1.0}$ & $0.4/1\mathrm{ x }10^{-12}$ & $-0.25/-1\mathrm{x}10^{-12}$ & $--$ & $--$ \\
\hline
%\multirow{4}{*}{\textbf{HG}} &
$\mathbf{2.0}$ & $0.7/1\mathrm{x}10^{-12}$ & $-2.4/-1\mathrm{x}10^{-12}$ & $1.3/0.93$ & $-4.4/-5.29$ \\
%\cline{2-4}
\hline
$\mathbf{3.0}$ & $0.95/1\mathrm{x}10^{-12}$ & $-7.0/-1\mathrm{x}10^{-12}$ & $1.9/1.47$ & $-9.1/-9.57$ \\
%\cline{2-4}
\hline
$\mathbf{4.0}$ & $1.2/2\mathrm{x}10^{-12}$ & $-12.8/-2\mathrm{x}10^{-12}$ & $2.4/2.25$ & $-14.7/-15.31$ \\
%\cline{2-4}
\hline
$\mathbf{5.0}$ & $1.6/2\mathrm{x}10^{-12}$ & $-19.4/-2\mathrm{x}10^{-12}$ & $2.95/2.89$ & $-21.05/-21.77$ \\
\hline
$\mathbf{6.0}$ & $2.1/3\mathrm{x}10^{-12}$ & $-26.8/-3\mathrm{x}10^{-12}$ & $--$ & $--$ \\
\hline
$\mathbf{7.0}$ & $2.7/16\mathrm{x}10^{-12}$ & $-34.7/-16\mathrm{x}10^{-12}$ & $--$ & $--$ \\
\hline
$\mathbf{8.0}$ & $3.3/30\mathrm{x}10^{-12}$ & $-43.2/-30\mathrm{x}10^{-12}$ & $--$ & $--$ \\
\hline
\end{tabular}
\caption{Critical currents $J_{z,crit}$ for skyrmion annihilation in 64 nm and 128 nm disks, at $T = $ 0 and 300 K. All current values are multiples of 10$^{12}$ A}.
\label{tab:crit_curr}
\end{table}
%%%%%%%%%%%%%%%%%%%%%%%%%%%%%%%%%%%%%%

%%%%%%%%%%%%%%%% Figure 13 %%%%%%%%%%%%%%%
\begin{figure}
\centering
\includegraphics[width=\textwidth]{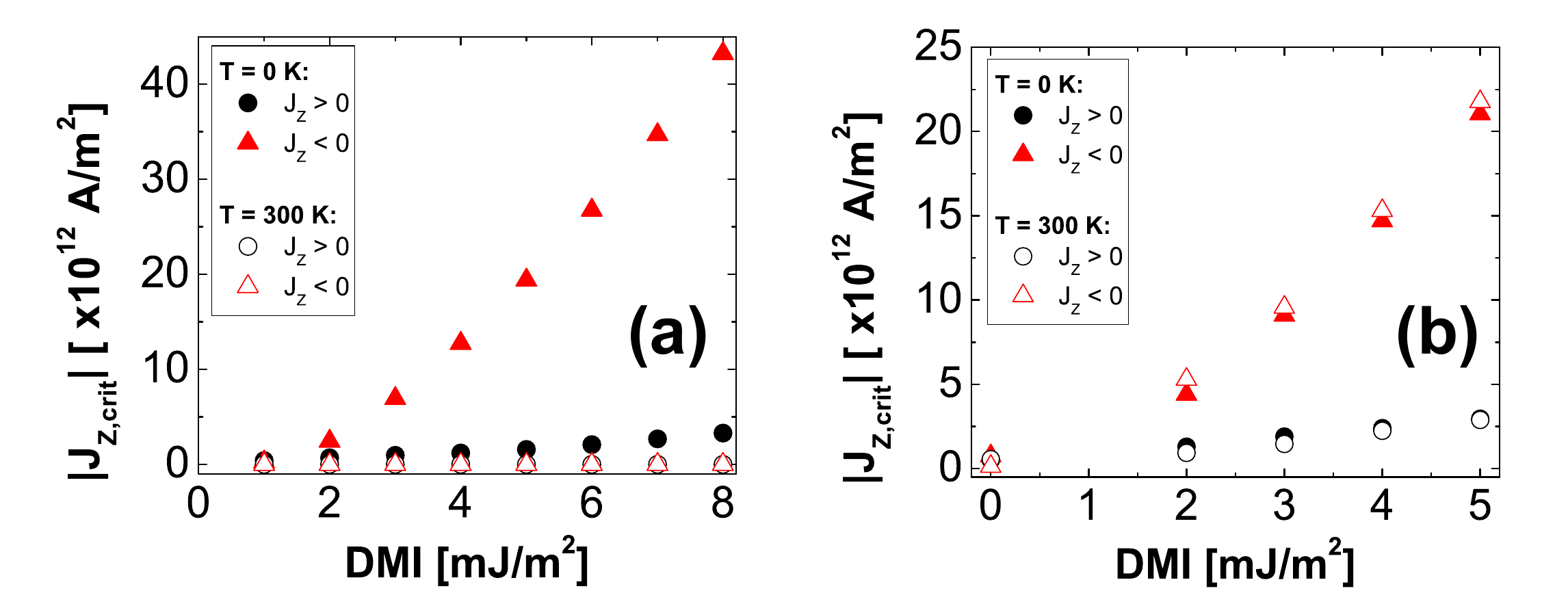}
\caption{Critical currents for spin polarized-induced skyrmion annihilation in function of DMI strength determined for (a) 64 and (b) 128 nm disks at $T = 0$ K (closed symbols) and $T = 300$ K (open symbols). (Color online). \label{fig:crit_curr_dmi}}
\end{figure}
%%%%%%%%%%%%%%%%%%%%%%%%%%%%%%%%%%%%%

\section{Conclusion \label{conclusion}}

Micromagnetic simulations of magnetic nanosize disks achieving their ground state from several initial configurations indicate that a single, confined skyrmion can be stable in these structures only for a few combinations of uniaxial anisotropy constant and Dzyaloshinskii-Moriya interaction, under ideal conditions ($T = 0$ K and defect-free sample). The results also show that this skyrmion ground state depends on the initial condition of the simulated disks. At room temperature, fewer of these combinations are available, as evidenced by the smaller $Sk = 1$ regions in the phase diagrams. This indicates that the fabrication of nanostructures capable of sustaining a stable skyrmion ground state relies critically on the fine tuning of these magnetic parameters especially for devices working at room temperature. 32 nm disks could only sustain a stable skyrmion a few cases, always at $T = 0$ K, while 64 and 128 nm disks had more combinations of $K_{u}$ and $D_{ex}$ for which a skyrmion ground state was stable, both at 0 K and 300 K. These ground states were always stable in the absence of magnetic fields, a fundamental condition for non-volatile data storage applications.

The results indicate that high values of uniaxial anisotropy and high values of DMI do not favor a skyrmion ground state in these magnetic disks. A high value of uniaxial anisotropy constant will not favor the smooth transition from $m_{z} = +1$ to $m_{z} = -1$ from the skyrmion center to its periphery, while a high value of DMI favors spin rotation leading to spirals or multidomain states. On the other hand, a small DMI is necessary to stabilize a single skyrmion in a disk, and in the absence of DMI a very fragile vortex-like skyrmion ground state is achieved only within a narrow range of $K_{u}$, for 128 nm, defect-free disks at both temperatures investigated. This skyrmion ground state is easily destroyed by magnetic fields or electric currents, but, surprisingly, it is robust against thermal agitation. The initial condition of the simulations is fundamental for the observation of these zero-DMI skyrmions, for even though pre-existent skyrmions can be stabilized even in the absence of DMI, the transition from a uniform magnetic state to a skyrmion is only observed in cases where a sizeable DMI is present. Furthermore, while all zero-DMI skyrmions observed were vortex-like, with the introduction of DMI causing them to become HG, a configuration favored by the interface-induced DMI employed in the simulations.

The skyrmion ground states observed can be quite stable under out-of-plane magnetic fields or spin-polarized currents. For positive fields/currents, the edge of the disk plays a major role in skyrmion annihilation, since the increase is its diameter eventually causes its border to touch the disk edge, leading to annihilation. For negative fields/currents, the critical fields/currents for skyrmion annihilation are much larger, because in this case their sense is antiparallel to the skyrmion core magnetization, effectively squeezing it and increasing the magnetostatic repulsion between the skyrmion walls. Finite temperatures introduce thermal agitation that tends to narrow stability regions in the phase diagrams, and to decrease the absolute values of the critical annihilation fields/currents. This is critical in the 64 nm disks, where it was observed that the skyrmions are easily destroyed by very small field/currents at room temperature. So, despite the larger skyrmion stability regions observed in their phase diagrams, 64 nm disks are less interesting for room temperature devices relying on magnetic fields or electrical currents for their operation. It is important to stress that in all these simulations defects in the samples were not considered. Structural defects commonly found in magnetic thin film-based nanostructures can introduce fluctuations in $K_{u}$ or $D_{ex}$, which lead to pinning and complex magnetization dynamics such as domain wall creep\cite{lemerle_prl,metaxas_prl,chauve_prb}. Defects are known to pin skyrmions in nanostructures\cite{boulle_nature}, but the effects of fluctuations in anisotropy or DMI on their formation and stability is still, to our knowledge, unknown, and could play a major role in their statics and dynamics. These results show that the conditions for skyrmion stability in a single magnetic disk demand a high degree of refinement in multilayer engineering and magnetic nanostructure fabrication, especially for room temperature applications, where achieving the right combination of $K_{u}$, DMI and diameter is paramount.
%%%%%%%%%%%%%%%%%%%%%%%%%%%%%%%%%%%%%%

%%%%%%%%%%% ACKNOWLEDGEMENTS %%%%%%%%%%%%%%%%%%
\begin{acknowledgements}
The authors are grateful to Dr. Alberto R. Torres for granting access to the computational facilities of the UFSC Physics Department. The authors acknowledge financial support from Brazilian agencies CNPq, CAPES and FAPERJ.
\end{acknowledgements}

%%%%%%%%%%%%%%%%% APPENDIX %%%%%%%%%%%%%%%%%%
%\appendix

%%%%%%%%%%%%%%%%%%%%%%%%%%%%%%%%%%%%%%%%
%\section{Connection between creep and hysteresis loops \label{app:creep}}

%%%%%%%%%%%%%%%%%%%%%%%%%%%%%%%%%%%%%%%%
\section*{References}
\bibliography{skyrmion_1_refs}

\end{document}